\documentclass[fleqn,usenatbib]{mnras}

\usepackage{newtxtext,newtxmath}

\usepackage[T1]{fontenc}

\usepackage{graphicx}	% Including figure files
\usepackage{amsmath}	% Advanced maths commands
\usepackage{gensymb}

\renewcommand{\today}{INR-TH-2023-010}

\usepackage{caption}
\usepackage{subcaption}

\newcommand{\black}{\color{black}}
\newcommand{\red}{\black}
\title[Neutrino flares of radio blazars]{Neutrino flares of radio blazars observed from TeV to PeV}

\author[A.~Suray \& S.~Troitsky]{
Alisa Suray$^{1}$\thanks{Corresponding author; e-mail: surai.ai21@physics.msu.ru}
and Sergey Troitsky$^{2,1}$
\\
$^{1}$Physics Department, Lomonosov Moscow State University, 1-2 Leninskie Gory,  Moscow 119991, Russia\\
$^{2}$Institute for Nuclear Research of the Russian Academy of Sciences, 60th October Anniversary prospect~7a, 117312 Moscow, Russia}

\date{Submitted to MNRAS Letters on June 29, 2023. Revised on August 30, 2023. Accepted on September 20, 2023}

\pubyear{2023}

\begin{document}
\label{firstpage}
\pagerange{\pageref{firstpage}--\pageref{lastpage}}
\maketitle

% Abstract of the paper
\begin{abstract}
Radio blazars have been linked both to individual high-energy neutrino events and to excesses in likelihood sky maps constructed from lower-energy neutrino data. However, the exact mechanism by which neutrinos are produced in these sources is still unknown. Here, we demonstrate that IceCube neutrinos with energies over 200~TeV, which were previously associated with bright radio blazars, are significantly more likely to be accompanied by flares of lower-energy events, compared to those lacking blazar counterparts. The parsec-scale core radio flux \red density \black of blazars, positioned within the error regions of energetic events, is strongly correlated with the likelihood of a day-scale lower-energy neutrino flare \red in directional and temporal coincidence with the high-energy event, \black reported by IceCube. The probability of a chance correlation is $3.6\times 10^{-4}$. This confirms the neutrino-blazar connection in a new and independent way, and provides valuable clues to understanding the origin of astrophysical neutrinos.
\end{abstract}

% Select between one and six entries from the list of approved keywords.
% Don't make up new ones.

%%%%%%%%%%%%%%%%%%%%%%%%%%%%%%%%%%%%%%%%%%%%%%%%%%

%%%%%%%%%%%%%%%%% BODY OF PAPER %%%%%%%%%%%%%%%%%%

\section{Introduction} 
\label{sec:intro}
High-energy neutrino sky is presently under active investigation, thanks to new data coming from the IceCube \citep{Icecube-Experiment}, ANTARES \citep{ANTARES-experiment}, Baikal-GVD \citep{Baikal-experiment} and KM3NeT \citep{KM3NeT-experiment} observatories. The presence of extraterrestrial neutrinos is well established by different experiments \citep{IceCubeFirst26,IceCube-HESE-2020,ANTARES2019ICRC,Baikal-diffuse}, while their origin is a matter of intense discussions, see e.g.\ \citet{Meszaros-rev,ST-UFN} for reviews. The neutrino sky is diverse: while a part of events probably originate in our Galaxy \citep{neutgalaxy,ANTARES-GalRidge,IceCube-Gal}, others have been associated with extragalactic sources, \citep[e.g.][etc.]{IceCubeTXSgamma,Resconi2020,neutradio1,Buson1,IceCube:NGC1068}. Evidence is growing that blazars, which are active galactic nuclei with relativistic jets pointing to the observer \citep{UrryPadovani-blazars}, are responsible for a significant part of the high-energy astrophysical neutrino flux. Some of the results refer to individual sources like TXS~0506$+$056 \citep{IceCubeTXSgamma,IceCubeTXSold,BaikalTXS} or others \citep{Kadler1424-418,Winter1502+106,Sahakyan0735+178}, but the most compelling evidence comes from statistical analyses of blazar populations versus neutrino data sets \citep[][etc.]{neutradio1,neutradio2,neutradio2022,Resconi2020,Hovatta2021,Franckowiak2022,Buson1,Buson2}.

High-energy neutrino telescopes work in a wide energy band, TeVs to PeVs. Some of previous studies considered highest-energy ($E \gtrsim 200$~TeV) events, selected individually as most probable candidates for astrophysical neutrinos. Other studies used entire data sets, dominated by lower-energy, $\sim$~TeV, events, most of which are caused by muons and neutrinos originating in the terrestrial atmosphere. Statistically significant associations with blazars have been reported both for the most energetic events \citep{Resconi2020,neutradio1,neutradio2022} and for excesses in the sky maps built from entire, mostly lower-energy data sets \citep{neutradio2,Buson1,Buson2}. This is remarkable since the efficient production of $\sim$~TeV and of $\sim10^2$~TeV neutrinos requires different conditions in the source, see Sec.~\ref{sec:disc:physics}. Because of the statistical nature of the most significant neutrino-blazar links, it was unclear if neutrinos of all energies from TeV to PeV are produced in the same class of sources, or in distinct blazar populations. In the former case, the question would arise whether high- and lower-energy neutrinos are produced simultaneously. The present work addresses these questions and searches for correlations of individual high-energy IceCube events, associated with blazars, with coincident flares of lower-energy neutrinos.

In the rest of the Letter, we discuss the data we use (Sec.~\ref{sec:data}), addressing separately the blazar sample (Sec.~\ref{sec:data:RFC}), high-energy neutrinos (Sec.~\ref{sec:data:alerts}) and lower-energy neutrino flares (Sec.~\ref{sec:data:flares}). In Sec.~\ref{sec:anal}, we perform the statistical analysis and present its results. We discuss the results and their implications in Sec.~\ref{sec:disc}. In particular, Sec.~\ref{sec:disc:blaz-all} stresses the importance of the selection of events associated with radio blazars for the result; Sec.~\ref{sec:disc:var-sets} discusses the potential impact of different IceCube event reconstructions on the blazar-related results. Implications of our results for modelling neutrino production in blazars are discussed in Sec.~\ref{sec:disc:physics}. We briefly conclude in Sec.~\ref{sec:concl}.

\section{Data} 
\label{sec:data}
The most direct observational feature of a blazar is its Doppler-boosted radiation from a spatially compact region, indicating a relativistic jet viewed at the angle of a few degrees. Localisation of the emission to parsec-scale cores of cosmologically distance sources is possible only with very-long-baseline interferometry (VLBI) in the radio band, due to its superb angular resolution. VLBI observations are an efficient tool to select blazars \citep[e.g.][]{BlandfordRev}, and a sample of VLBI-selected radio blazars was used to establish, in a statistical way, associations of these sources both with highest-energy \citep{neutradio1,neutradio2022} and lower-energy \citep{neutradio2} neutrinos. In the present work, we use the same samples of energetic neutrinos and of radio blazars from \citet{neutradio2022}, supplemented by recently published results of IceCube searches for associated lower-energy flares.

\subsection{IceCube high-energy events} 
\label{sec:data:alerts}
Quality cuts for the highest-energy neutrino sample were fixed by \citet{neutradio1}, motivated by requirements of good reconstruction accuracy and of high ratio of the astrophysical signal to the atmospheric background. The cuts select, from event lists published by IceCube, muon tracks with expected progenitor neutrino energy $E_\nu > 200$~TeV and the solid angle enclosed by the 90\% CL directional uncertainty contour in the sky $\Omega_{90\%}<10~\mbox{deg}^2$.
The same, up to addition of more recent events satisfying the same selection criteria, sample was used in a number of subsequent studies \citep{Hovatta2021,neutradio2022,neutxray,neutgalaxy}. Here, we start with the sample of 71 events \citep{neutradio2022} supplemented by two recent IceCube alerts issued later \citep{221223A-obs,221229A-obs} and passing the same selection cuts. We have to remove 12 events for which no flare data, discussed in Sec.~\ref{sec:data:flares}, are available (in particular, 10 of them were registered before May 2011). Finally, we remove one event for which \citet{IceCat-1} reported a coincident IceTop signal, strongly suggesting that this event was atmospheric. The list of 60 events in the final high-energy sample is presented in Supplementary Information~I. We recall that about $1/3$ of these events are still expected to be of the atmospheric origin \citep{neutradio1,neutgalaxy}, but there is no direct way to identify, which ones.
In what follows, we call all these 60 events ``HE neutrinos'' for brevity, though some of them may be caused by atmospheric muons and not neutrinos.

\subsection{IceCube lower-energy flares} 
\label{sec:data:flares}
For each of the public high-energy neutrino alert, IceCube distributes also the information about accompanying lower-energy neutrinos. For archival events between May 2011 and December 2020, this information was collected and published by \citet{IceCube-flares}. For events of 2021--2022, we use the data from GCN circulars \citep{210210A-fl,210811A-fl,210922A-fl,220205B-fl,220303A-fl,220306A-fl,220425A-fl,220513A-fl,221223A-fl,221229A-fl}. 

These publications present high-level data for each energetic event, which include the p-value $p$, measuring the probability that energies and arrival directions of other neutrinos, detected within a certain time window $\pm T$ around the HE neutrino arrival moment, are consistent with the background. Details of the calculation of $p$ are presented by \citet{IceCube-flares}. In our analysis, we use $L \equiv -\log_{10} p$ as a convenient variable (our statistical procedure is not sensitive to any monotonic redefinition of the observable) and call it ``likelihood of a flare'' for brevity. Of two values of $T$ for which $p$ is reported, 500~s and 1~day, we choose the latter because longer flare durations are suggested by multimessenger observations \citep{IceCubeTXSgamma,neutradio1,Hovatta2021,BaikalTXS}. Note that for any individual event, the accompanying flare is not significant, with the highest $L=1.52$, corresponding to a $2.2\sigma$ excess.

\subsection{Radio blazars} 
\label{sec:data:RFC}
Following previous works, we use a complete full-sky flux-density limited sample of blazars selected from the Radio Fundamental Catalog\footnote{\url{http://astrogeo.org/sol/rfc/rfc_2022b/}. Note that we use the RFC 2022b version for full consistency with \citet{neutradio2022}, though a newer version is available.} (RFC) by requiring the 8-GHz VLBI average flux density to exceed 0.15~Jy. The RFC sample was compiled from dedicated observations by \citet{2002ApJS..141...13B,2003AJ....126.2562F,2005AJ....129.1163P,2006AJ....131.1872P,2008AJ....136..580P,2009JGeod..83..859P,2011AJ....142...35P,2011MNRAS.414.2528P,2019MNRAS.485...88P,2011AJ....142..105P,2012MNRAS.419.1097P,2013AJ....146....5P,r:wfcs,2007AJ....133.1236K,2012A&A...544A..34P,2012ApJ...758...84P,2015ApJS..217....4S,2016AJ....151..154G,2017ApJS..230...13S}. The isotropy of the sample of potential sources is important when other, anisotropic astrophysical contributions to the total neutrino flux may be present \citep{neutgalaxy}, e.g.\ the Galactic one. In the present work, this source sample is used to associate radio blazars with simulated neutrino events at the stage of Monte-Carlo estimation of the statistical significance (Sec.~\ref{sec:anal}) and to search for blazar counterparts of two most recent HE neutrinos, see Sec.~\ref{sec:data:alerts}. For the rest of events, counterparts \red from the same RFC~2022b catalog \black were determined by \citet{neutradio2022}.

\section{Analysis and results}
\label{sec:anal}
We address a question, whether blazar-associated HE neutrino events are accompanied by lower-energy neutrino flares. To this end, we obtain, for each HE neutrino, two quantities: one is related to the probability of an association with a blazar, another measures the probability of a neutrino flare. Then the correlation between these two quantities is searched for by standard statistical methods. To estimate how often this or stronger correlation occurs by chance, we use Monte-Carlo simulations. 

\begin{figure}
\centering
\includegraphics[width=0.95\linewidth,trim=160 200 305 120,clip]{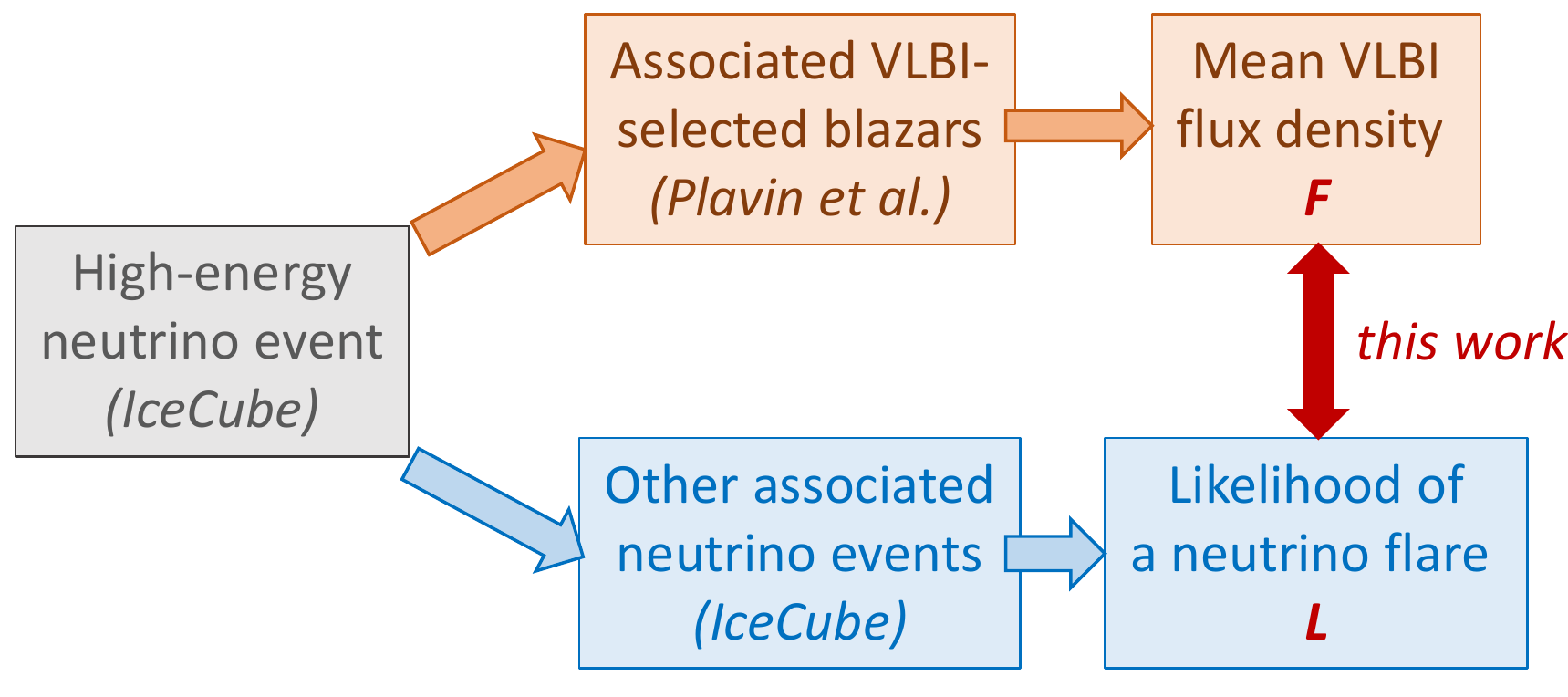}
\caption{
\label{fig:scheme}
Scheme of the analysis of a high-energy event.
}
\end{figure}

For an individual HE neutrino, we use the flare likelihood $L$, published by IceCube and discussed in Sec.~\ref{sec:data:flares}, as the measure of the probability of an associated neutrino flare. To quantify associations with blazars, we follow \citet{neutradio1,neutradio2022} and use the mean 8-GHz VLBI flux density $F$ of blazars from the RFC complete, $F>0.15$~Jy, sample, associated with the events. If more than one blazar is associated with a HE neutrino, $F$ is taken to be equal to their mean flux, while $F=0$ is set if no associated blazars are present. We use the same \red procedure \black for association as determined by \citet{neutradio1,neutradio2022}, and recollect all the details of this procedure in Supplementary Information~II for reference. Therefore, both quantities, $L$ and $F$, are determined in previous works, see Fig.~\ref{fig:scheme}.

\begin{figure}
\includegraphics[width=0.95\linewidth]{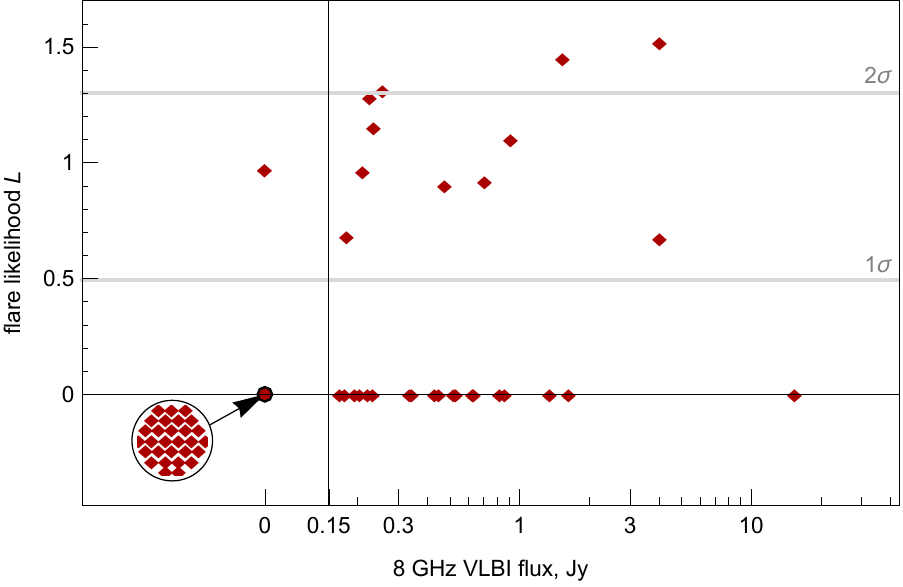}
\caption{
\label{f:listplot}
The likelihood $L$ of a lower-energy neutrino flare coincident in time and direction with a high-energy event versus the mean \red logariphmic \black 8-GHz VLBI flux density $F$ of blazars associated with the same event. The vertical line indicates the 0.15~Jy flux-density threshold of the sample. Zero flux means no associated blazars, $L=0$ means no neutrino \red flare \black. The circular inset illustrates 29 \red HE \black events without associated blazars and  neutrino \red flares\black. Of 12 events with neutrino \red flares\black, 11 are associated with blazars (92\%), while only 19 of 48 (40\%) of events without \red flares \black have blazar counterparts. 
}
\end{figure}

We now have two sets, $\{F_i\}$ and $\{L_i\}$, where $i=1,\dots, 60$ enumerates HE neutrinos. The values of $F_i$ and $L_i$ are collected in Table~\ref{t:events} in Supplementary Information~I and presented graphically in Fig.~\ref{f:listplot}.
We want to test the null hypothesis that the association of HE neutrinos with radio blazars is unrelated to the low-energy neutrino flares, that is the sets $\{F_i\}$ and $\{L_i\}$ are uncorrelated. To do this, we introduce test statistics which measure the correlation. We then perform Monte-Carlo simulations to see how strong the correlation is.

The relation between the value of $L> 0$ and properties of the possible neutrino flare is unknown, but $L=0$ corresponds to the absence of accompanying \red low-energy \black neutrinos. This motivates the simplest statistics equal to the ratio of mean flux densities of blazars associated with HE neutrinos with $L\ne 0$ to those with $L=0$, the flux ratio statistics $f$. We also consider two more common test statistics for correlations between two sets \red of numbers\black, the Spearman rank statistics $\rho$ and the Kendall rank correlation coefficient $\tau$. For reference, we collect explicit definitions of $f$, $\rho$ and $\tau$ in Supplementary Information~III.

Though some expressions exist which relate values of $\rho$ or $\tau$ to the probability of chance correlations, they requre certain assumptions about the distribution of uncorrelated values. Therefore we do not use these relations and perform Monte-Carlo simulations to assess the significance of correlations. We simulate artificial HE neutrino sets by assigning random right ascensions to real \red HE \black events, keeping declinations, shapes and sizes of error regions, and values of $L$ unchanged. This approach is justified because the sensitivity of IceCube depends on the zenith angle only; any potential dependence on the azimuthal angle related to the detector geometry is washed out by the Earth's rotation. Scrambling of right ascensions is the standard way to randomize arrival directions in IceCube anisotropy studies \citep[e.g.,][]{IceCube-new-blazars}. 

For each artificial set of 60 \red HE \black neutrinos, we find associated blazars in the same way it was done for the real data, find the values of $F$ and calculate the same test statistics. We repeat the simulations $10^5$ times. The probability of chance correlation is given by the fraction of simulated sets with the same or larger value of the test statistics as obtained for the real data. The results are presented in Table~\ref{t:results}.
\begin{table}
\centering
\begin{tabular}{ccc}
\hline\hline
Test &  Observed & Chance\\
statistics &  value & probability\\
\hline
$f$   &  2.202& $3.0\times 10^{-4}$\\
$\rho$&  0.419& $3.6\times 10^{-4}$\\
$\tau$&  0.363 & $3.5\times 10^{-4}$\\
\hline
\end{tabular}
\caption{\label{t:results}
Results for three test statistics (see the text for details).  
}
\end{table}

\begin{figure}
\includegraphics[width=0.95\linewidth]{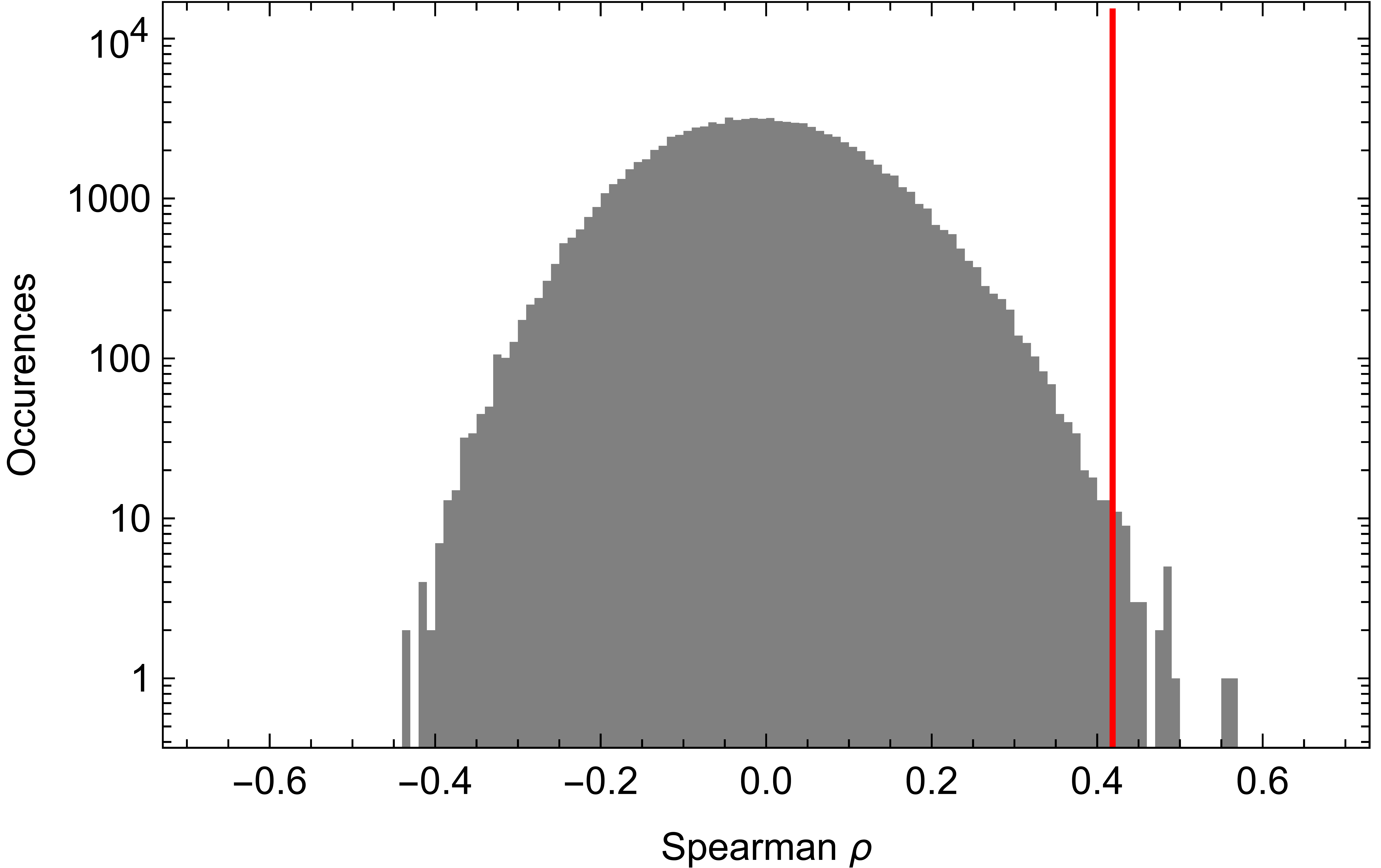}
\caption{
\label{f:spearman}
Distribution of Spearman rank test statistics $\rho$ for $10^5$ randomized data sets (gray histogram) and the value of $\rho_0$ for the real data (vertical red line). Only 36 of $10^5$ simulated sets have $\rho \ge \rho_0$.
}
\end{figure}

We see that all three tests give similarly low chance probabilities of correlations and conclude that the HE neutrinos associated with radio blazars are significantly more likely accompanied by lower-energy neutrino flares, compared to HE neutrinos lacking blazar counterparts. To be conservative with respect to the three trials $(f,\rho,\tau)$, we choose to report the highest chance probability (the lowest significance) as our final result. This is given by the Spearman rank test and corresponds to the equivalent Gaussian significance of $3.6\sigma$. The distribution of simulated values of $\rho$ is presented in Fig.~\ref{f:spearman}.
Note that the distribution is slightly asymmetric, which confirms the need of Monte-Carlo simulations to assess the strength of correlations. Similar plots for $f$ and $\tau$ are presented in Supplementary Information~III.

\section{Discussion}
\label{sec:disc}

\subsection{Blazar-associated alerts versus all alerts}
\label{sec:disc:blaz-all}
In a dedicated study, \citet{IceCube-flares} addressed a similar question, whether high-energy events are accompanied by lower-energy neutrino flares. They considered all events from the catalog by \citet{IceCat-1} and found that the distribution of $L$ for all events is consistent with that expected for random coincidences. \red We cannot test this result for our sample of 60 events, selected in a different way, because this would require unpublished information. However, \black  we demonstrate that a significant correlation exists between $L$ and the VLBI flux \red density \black of associated blazars. This suggests that the association with blazars changes the situation drastically. Among other implications, this gives a strong independent support to the original association \citep{neutradio1,neutradio2022} of HE neutrinos with radio blazars. 

\red Let us estimate roughly the fraction of HE neutrinos coming from radio blazars. \black The events without a blazar association, or those randomly coincident with blazars, are not expected to correlate with short-time neutrino flares. About $1/3$ of the HE neutrinos in the sample, that is $\sim 20$ of 60, are atmospheric \citep{neutradio2022}. Of the $\sim 40$ \red remaining \black astrophysical events, $(28 \pm 9)\%$, that is $\sim(11 \pm 4)$, are expected to have a Galactic origin \citep{neutgalaxy}. The original work by \citet{neutradio1} estimated the fraction of HE neutrinos physically associated with blazars from the RFC complete sample as $\sim 10/35 \approx 28.5\%$. \red \citet{neutradio2} estimated that $\sim 25\%$ of lower-energy neutrinos come from radio bright blazars from the same sample. This is in formal agreement with the 90\% CL upper bound by \citet{Zhou}, $\lesssim 30\%$, though both estimates were obtained in different frameworks and have their own systematics. Altogether, \black we expect that $\sim 11$ HE neutrinos out of 60 are related to blazars from the catalog we analyze. \red This matches the actual numbers: \black 12 HE events out of 60 have nonzero flare likelihood $L$, 11 of which have blazar counterparts. \red Given the small statistics and large fluctuations, one generally does not expect that exact coincidence of numbers, which however demonstrates the overall consistency of estimates. \black 

\red
Note that the correlation we observe is not dominated by any single source. In particular, this relates to TXS~0506$+$056, established as a neutrino source independently of, and before, the radio-blazar statistical studies. Contrary, the event of 2017-09-22, associated with this blazar, was not accompanied by a \emph{simultaneous} low-energy neutrino flare. The largest contribution to the correlation we find here comes from PKS~1731$-$038, associated with two HE neutrinos, both accompanied by lower-energy flares. It was claimed as a probable neutrino source in the first study of radio blazars and HE neutrinos by \citet{neutradio1}, based on the event of 2011-09-30. Subsequently, the second HE neutrino arrived from its direction on 2022-02-05 \citep{PKS1741ATel,neutradio2022}. Since it was claimed as a neutrino source in the radio-blazar analysis, and not elsewhere, dropping it from the sample would not be justified. However, we checked that even without the two events associated with PKS~1731$-$038, our result remains pronounced ($3.0\sigma$). 

PKS~1731$-$038 is among the brightest radio blazars in the sky and the only blazar associated with two HE neutrinos in the sample we use. It also dominates the correlation between HE neutrinos and hard X-ray blazars reported by \citet{neutxray}. The present study further motivates this individual association.
\black

Due to a statistical nature of the study, one cannot be sure that all of the corresponding blasars are the neutrino sources. The remaining part of \red extragalactic \black HE neutrinos may include events coming from blazars beyond the completeness limit of the RFC catalog or from other extragalactic sources like Seyfert galaxies \citep{IceCube:NGC1068,SeyfertSemikoz}, tidal disruption events \citep{TDE-2005.05340, TDE-2111.09390}, etc.

\subsection{Radio-blazar associations in various IceCube reconstructions}
\label{sec:disc:var-sets}
Arrival directions of individual neutrinos differ between various published IceCube data sets, indicating the presence of systematic uncertainties in reconstruction \citep[see e.g.][and references therein]{IceCubeSyst,ST-UFN}. This is expected given the lack of precise knowledge of ice properties through the km$^3$ detector volume. In addition, the energies of neutrinos detected by muon tracks, used in our sample, are determined with huge statistical uncertainties, reaching an order of magnitude, see e.g.\ \citet{IceCubeTXSgamma}. Previous studies indicated that blazar associations, found with the help of published IceCube data, may become less pronounced when data obtained in a different reconstruction are used. For instance, the IceCube 7-year sky map \citep{IceCube7yrmap-data} was used to establish neutrino-blazar associations \citep{neutradio2,Buson1}, but the 10-year public event catalog \citep{IceCube10yrcat-arXiv} did not reveal a significant signal (\citealt{Zhou}; though the results of the two analyses do not contradict to each other, \citealt{PlavinICRC2021}). At energies above 200~TeV, the sample used by \citet{neutradio2022} (and here) collected event details as they were originally published, and demonstrated significant correlations with radio blazars, but in a similar analysis by \citet{IceCube-new-blazars}, based on recently reprocessed IceCube data \citep{IceCat-1}, the effect was found to be much weaker. We repeated our present analysis with the catalog of \citet{IceCat-1} and confirmed that the radio-blazar correlation is not pronounced in this case. Consequently, the correlation of blazar-associated energetic events with lower-energy neutrino flares does not show up for the \citet{IceCat-1} alert-like event list.

Understanding the reasons for the discrepancies would require a detailed comparison of the data sets and reconstruction procedures, hardly possible outside the IceCube collaboration. Associations of $E>200$~TeV neutrinos with VLBI-selected blazars, found by \citet{neutradio1}, strengthened (from $3.1\sigma$ to $3.6\sigma$) with the addition of the new data \citep{neutradio2022}, and our present study adds confidence to \red the origin of a significant part of HE neutrinos in radio blazars \citep{neutradio1}\black. One possible reason for weakening the neutrino-blazar signal in the reconstruction of \citet{IceCat-1} may be related to significant changes in the estimated energies of events. This may or may not be related to the extensive use of machine learning in newer reconstructions \citep{IceCat-1}, which reduces statistical uncertainties but relies heavily on the simulated training sets, for which certain ice properties should be assumed. Future studies with new liquid-water detectors, Baikal-GVD  \citep{Baikal-experiment} and KM3NeT \citep{KM3NeT-experiment}, are expected to shed light on these issues.

\subsection{Implications for neutrino production mechanisms}
\label{sec:disc:physics}
Theoretical interpretation of the observational results of the present work will be addressed elsewhere, but some general implications should be mentioned.

Firstly, our result implies that the same blazars, which produce neutrinos of very high, $E>200$~TeV, energies, are capable to produce simultaneously lower-energy neutrinos. This is nontrivial, because in blazars, neutrinos are expected to be produced in proton-photon collisions \citep[e.g.][]{Boettcher-rev}, and the cross section of this process is dominated by the $\Delta^+$ resonance peak. This means that for efficient production of neutrinos, say, of $\sim 10$~TeV and of $\sim 300$~TeV, one needs to have target photons with energies 30 times different \citep[see e.g.][for discussion and more references]{neutxray}. Therefore, narrow thermal spectra of target photons are disfavoured. Production of lower-energy neutrinos requires target photons of proportionally higher energies, reaching the hard X-ray band. These hard X rays, probably Doppler-boosted, are indeed observed from the blazars associated with $E>200$~TeV events \citep{neutxray}, in agreement with the present study. This reasoning does not apply to exotic proton-proton mechanisms \citep{Neronov:JETPL2020} which however explain only sub-PeV neutrinos from blazars and do not address lower energies. More detailed studies of yet unpublished and future data would help to better understand the neutrino spectra of blazars and to constrain the neutrino production mechanisms.

Secondly, the timescale of the flares we studied, determined by the availability of data, was very short, $\pm 1$~day. An excess of the neutrino flux at longer intervals around the arrival of HE neutrinos,  not studied by \citet{IceCube-flares} but covering this 2-day interval, is not excluded. In fact, various observational studies suggest these longer intervals for blazar neutrino flares \citep{IceCubeTXSold,neutradio1,Hovatta2021}. This matches the expectations of models in which relevant proton-photon interactions take place at parsec-scale distances from the central black hole, and the target photons have Compton origin \citep{neutradio2,Kivokurtseva}. Studies of longer neutrino flares are thus welcome. However, even day-scale neutrino flares might be explained in a two-zone model by \citet{Kivokurtseva}, because the relevant variability may be inherited from the proton acceleration mechanism, working close to the black hole.

\section{Conclusions}
\label{sec:concl}
Our study demonstrates that energetic neutrino events coming from radio blazars are accompanied by \red simultaneous day-scale \black lower-energy neutrino flares from the same directions. The statistical significance of the correlation exceeds $3.6\sigma$ (Gaussian equivalent). Contrary, no significant correlation were found for all events, without a selection by a blazar association. This means that neutrinos of energies from a few TeV to several hundred TeV are jointly produced in radio blazars. This result adds more confidence to previously established association of radio blazars and neutrinos, confirming that about $25\%$ of the high-energy astrophysical neutrino flux originate from the VLBI-selected blazars. In the frameworks of the baseline proton-photon mechanism, our observation suggests the presence of a wide spectrum of target photons in the neutrino production zone. Future high-quality data from neutrino telescopes will help to understand the mechanism of neutrino production in blazars.

\section*{Acknowledgements}
We thank Alexander Plavin and Grigory Rubtsov for helpful discussions of the statistical analysis and of implications of this work. \red We are indebted to Yuri Kovalev and to the anonymous reviewer for careful reading of the manuscript and for useful comments. \black

\section*{Data Availability}
The analysis is entirely based on published data, referenced in the text.
%the IceCube neutrino data, collected from public sources as described by \citep{neutradio3} and supplemented by the data from Table~ of \citep{IceCube-flares}. Radio VLBI observations compiled in the Astrogeo\footnote{\url{ http://astrogeo.org/vlbi_images/}} database and the Radio Fundamental Catalogue\footnote{\url{http://astrogeo.org/sol/rfc/rfc_2022b/}}. The IceCube neutrino detections are collected from the published sources as described in \autoref{s:data_icecube}.

\bibliographystyle{mnras}
\bibliography{neutrino-flares}
\label{lastpage}

\clearpage
%\appendix
\section*{Supplementary information I:\\
List of events}
\label{a:events} 
Table~\ref{t:events} presents the list of 60 high-energy IceCube events used in the present work. Additional information on the events may be found in \citet{neutradio2022} using event dates (Col.~1) as ID, with the exception of two last events for which we used GCN \citep{221223A-obs,221229A-obs} alert information. Column~2 gives 8-GHz VLBI fluxes of associated blazars (see Supplementary Information~II for details; mean if several blazars are associated; zero if no blazar from the RFC complete sample is associated). Column~3 gives the values of flare likelihood $L$, discussed in Sec.~\ref{sec:data:flares}.
\begin{table}
\centering
\begin{tabular}{ccc}
\hline\hline
Event date & VLBI flux, Jy & $L$\\
\hline
2011-06-10   &   0.    &    0.   \\
2011-07-14   &  1.358  &   0.    \\
2011-09-30   &  4.034  &   0.67  \\
2012-03-01   &   0.    &    0.   \\
2012-05-15   &  0.916  &   1.1   \\
2012-05-23   &  0.821  &   0.    \\
2012-08-07   &   0.    &    0.   \\
2012-09-22   &   0.    &    0.   \\
2012-10-11   &  0.225  &   1.28  \\
2012-10-26   &   0.    &    0.   \\
2013-06-27   &  0.337  &   0.    \\
2013-10-14   &  0.628  &   0.    \\
2013-10-23   &  0.176  &   0.    \\
2013-12-04   &  0.236  &   1.15  \\
2014-01-08   &  0.179  &   0.68  \\
2014-01-09   &   0.    &    0.   \\
2014-01-22   &   0.    &    0.   \\
2014-02-03   &  0.519  &   0.    \\
2014-06-09   &   0.    &    0.   \\
2014-06-11   &   0.    &    0.   \\
2014-09-23   &   0.    &    0.   \\
2015-01-27   &   0.    &    0.   \\
2015-05-15   &   0.    &    0.   \\
2015-07-14   &   0.    &    0.   \\
2015-08-12   &  1.532  &   1.45  \\
2015-08-31   &  1.627  &   0.    \\
2015-09-04   &  0.341  &   0.    \\
2015-09-23   &   0.    &    0.   \\
2015-09-26   &   15.383&    0.   \\
2015-11-14   &  0.636  &   0.    \\
2015-11-22   &   0.    &    0.   \\
2016-01-28   &  0.865  &   0.    \\
2016-03-31   &  0.205  &   0.    \\
2016-05-10   &   0.    &    0.   \\
2016-07-31   &   0.    &    0.   \\
2016-08-06   &   0.    &    0.   \\
2016-12-10   &   0.    &    0.97 \\
2017-03-21   &   0.429 &    0.   \\
2017-09-22   &   0.447 &    0.   \\
2017-11-06   &   0.211 &    0.96 \\
2018-09-08   &   0.258 &    1.31 \\
2018-10-23   &   0.    &    0.   \\
2019-05-03   &  0.     &   0.    \\
2019-07-30   &  0.53   &   0.    \\
2020-06-15   &   0.    &    0.   \\
2020-09-26   &   0.    &    0.   \\
2020-10-07   &   0.    &    0.   \\
2020-11-14   &   0.    &    0.   \\
2020-11-30   &  0.476  &   0.9   \\
2020-12-09   &   0.    &    0.   \\
2021-02-10   &   0.    &    0.   \\
2021-08-11   &  0.194  &   0.    \\
2021-09-22   &   0.    &    0.   \\
2022-02-05   &  4.034  &   1.52  \\
2022-03-03   &  0.71   &   0.92  \\
2022-03-06   &   0.    &    0.   \\
2022-04-25   &  0.168  &   0.    \\
2022-05-13   &  0.221  &   0.    \\
2022-12-23   &   0.    &    0.   \\
2022-12-29   &  0.232  &   0.    \\
\hline
\end{tabular}
\caption{\label{t:events}
Data used in the present study (see the text for details).  
}
\end{table}

\section*{Supplementary information II:\\
Procedure to calculate the associated blazar flux $F$}
\label{a:error-region}
To determine $F$, one needs to determine which blazar is associated with a particular HE neutrino. We follow the procedure defined by \citet{neutradio1}. Details of the procedure have been fixed in previous works and extensively discussed there. Here we keep the procedure unchanged and describe it for reference.

\textbf{1.} The 90\% CL uncertainty in the arrival direction of a HE neutrino is reported by IceCube as
\begin{equation}
\left( 
\alpha^{+\Delta \alpha_+}_{-\Delta \alpha_-},  
\delta^{+\Delta \delta_+}_{-\Delta \delta_-} 
\right), 
    \label{eq:plus-minus}
\end{equation}
where $\alpha$ and $\delta$ are equatorial coordinates and $\Delta \alpha_\pm$, $\Delta\delta_\pm$ are their two-sided uncertainties. Define the 90\% CL containment region in the sky enclosed by four quarters of ellipses with corresponding half-axes of $1.3\Delta \alpha_\pm$, $1.3\Delta\delta_\pm$, where the coefficient 1.3 comes from the difference in definitions between one-dimensional and two-dimensional errors. 

\textbf{2. } Enlarge the obtained error region by $\Delta_s=0.45^\circ$ in all directions to compensate for unaccounted systematic errors. The value of $\Delta_s$ has been optimized previously, and results of previous works have taken into account the look-elsewhere effect related to this optimization. Here we fix the value of $\Delta_s=0.45^\circ$ determined previously, do not optimize it and therefore do not need to introduce the corresponding trial correction to our statistical results. Note that the optimal value of $0.5^\circ$ in \citet{neutradio1} was found with a computer code having a minor inconsistency \citep{neutradio2022}, and the correct value of $0.45^\circ$ applies to both the original \citep{neutradio1} and updated \citep{neutradio2022} samples. Note also that \citet{IceCube-new-blazars} revealed that, at least for recent events, the 90\% CL uncertainties reported in the form (\ref{eq:plus-minus}) by IceCube do not have the meaning of one-dimensional statistical errors, but indicate the corners of a rectangular region in the sky containing the 90\% uncertainty area, hence the multiplication by 1.3 is not needed. \citet{neutxray} demonstrated that without the multiplication, the optimal value of $\Delta_s$ becomes equal to 0.78, without any change in results with respect to the previously defined procedure. In the present work, we use the established procedure of multiplying by 1.3 and adding $\Delta_s=0.45^\circ$.

\textbf{3. } A source from the complete VLBI-selected RFC sample discussed in Sec.~\ref{sec:data:RFC}, which falls within the enlarged error contour described above, is considered as associated with the corresponding HE neutrino. If one source is associated, then $F$ equals to its 8-GHz VLBI flux density from the RFC catalog. In case there are several sources associated with one HE neutrino, $F$ equals to the geometric mean of their flux densities. If no blazar is associated, then $F=0$. Note that the RFC catalog gives the time-averaged historical flux densities and hence does not directly trace possible flares at the HE neutrino arrival time.

\section*{Supplementary information III:\\
Test statistics for correlations}
\label{a:statdetails}

\subsection*{III.1. The flux-ratio $f$ statistics}
\begin{figure}
\includegraphics[width=0.95\linewidth]{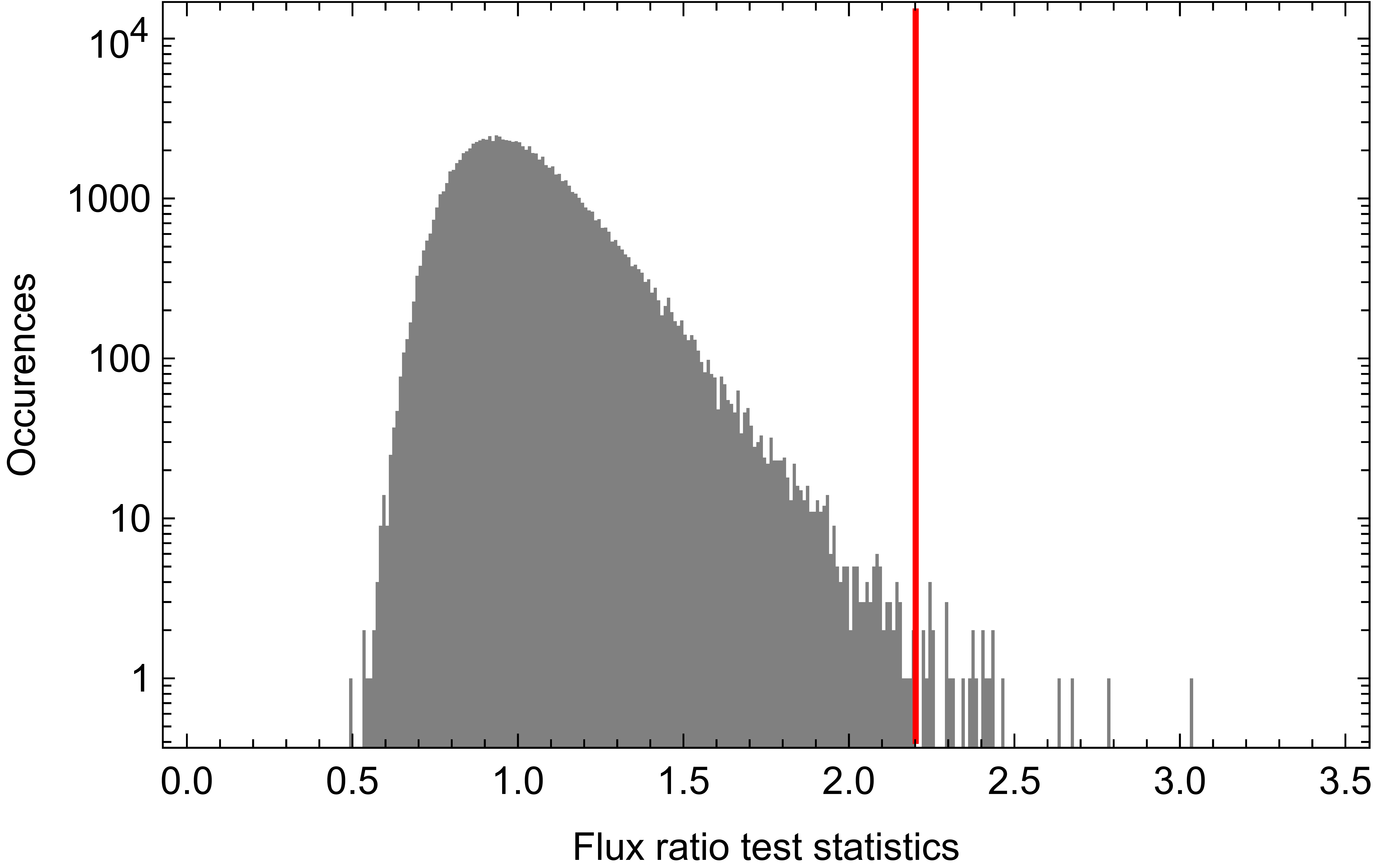}
\caption{
\label{fig:fluxrat}
Distribution of the flux-ratio test statistics $f$ for $10^5$ randomized data sets (gray histogram) and the value of $f_0$ for the real data (vertical red line). Only 30 of $10^5$ simulated sets have $f \ge f_0$.
}
\end{figure}

The flux-ratio test is a simple ``yes or no'' test motivated for the present study because a relation between the likelihood value $L$ and the corresponding flare flux is not known. It measures the deviation from a random expectation in fluxes of blazars associated with events with nonzero $L$, compared to those with $L=0$. We define the corresponding test statistics $f$ below and use Monte-Carlo simulations, Sec.~\ref{sec:anal}, to relate the value of $\tau$ to the significance of correlations, see Fig.~\ref{fig:fluxrat}.

Consider the set $\{F_i\}$ of $n=60$ mean blazar fluxes and the set $\{L_i\}$ of corresponding flare likelihoods. The flux ratio statistics is determined as the ratio of mean fluxes,
$$
f=\frac{\left\langle F_{i:\, L_i \ne 0}\right\rangle}{\left\langle F_{i:\, L_i = 0}\right\rangle}.
$$
Like elsewhere in the paper, the mean flux is understood as the geometric mean.

\subsection*{III.2. The Spearman $\rho$ statistics}
The Spearman rank test is a standard tool for estimation of the significance of correlations between two sets of data. For reference, we present here the definition of the corresponding test statistics $\rho$. We use Monte-Carlo simulations, Sec.~\ref{sec:anal}, to relate the value of $\rho$ to the significance of correlations, see Fig.~\ref{f:spearman} in Sec.~\ref{sec:anal}. 

Consider the set $\{F_i\}$ of $n=60$ mean blazar fluxes and the set $\{L_i\}$ of corresponding flare likelihoods. For each $L_i$, denote as $r_{L,i}$ its rank (the number it would have in an ordered list of the same values). Some of values may repeat (e.g., a large number of zero values); the ranks of corresponding elements are then replaced by their average, so that all equal values in the list have equal ranks (possibly non-integer). Let $m_{L,j}$ be the multiplicity of each value of $r_{L,i}$, where $j$ runs over non-equal values of ranks in $\{L_i\}$.

The list $\{F_i\}$ also has equal elements, including a number of zero elements and those coming from two events associated with the same blazar, PKS 1741$-$038. Let $r_{F,i}$ and $m_{F,k}$ be the ranks and multiplicities defined in a similar way, where $k$ runs over non-equal values of ranks in $\{F_i\}$. The test statistics is then
$$
\rho=\frac{\frac{1}{6}\left(n^3-n\right) - T_F - T_L - \sum_i\left(r_{F,i}-r_{L,i}\right)^2}{\sqrt{ \left( \frac{1}{6}\left(n^3-n\right) - 2 T_F\right) \left(\frac{1}{6}\left(n^3-n\right) - 2 T_L \right)}},
$$
where the corrections for ties are
$$
T_F=\frac{1}{12}\sum_k \left(m_{F,k}^3-m_{F,k} \right),
$$
$$
T_L=\frac{1}{12}\sum_j \left(m_{L,j}^3-m_{L,j} \right).
$$
\subsection*{III.3. The Kendall $\tau$ statistics}
\begin{figure}
\includegraphics[width=0.95\linewidth]{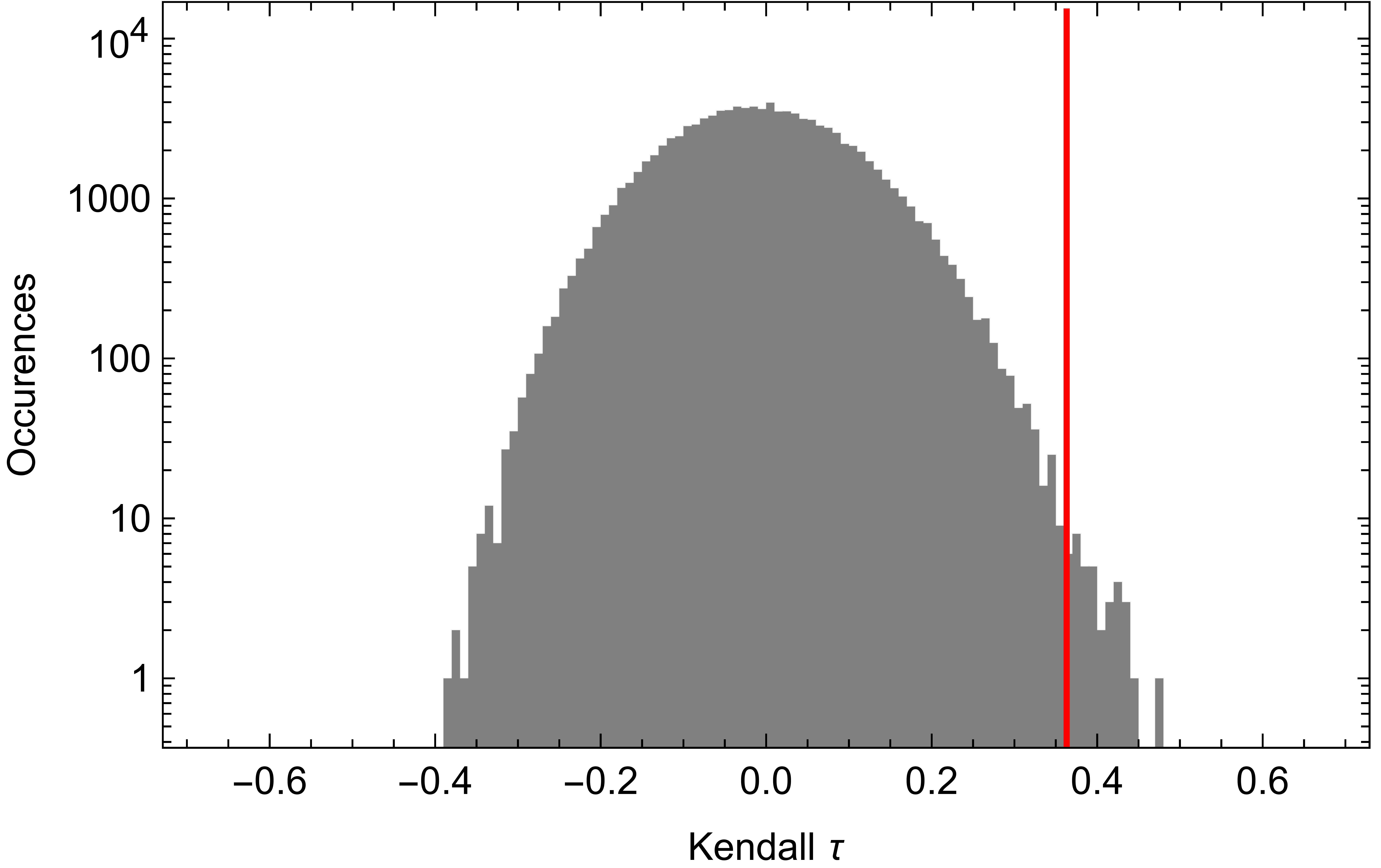}
\caption{
\label{fig:Kendall}
Distribution of Kendall test statistics $\tau$ for $10^5$ randomized data sets (gray histogram) and the value of $\tau_0$ for the real data (vertical red line). Only 35 of $10^5$ simulated sets have $\tau \ge \tau_0$.
}
\end{figure}

The Kendall rank correlation coefficient is another frequently used tool for estimation of the significance of correlations between two sets of data. For reference, we present here the definition of the corresponding test statistics $\tau$. We use Monte-Carlo simulations, Sec.~\ref{sec:anal}, to relate the value of $\tau$ to the significance of correlations, see Fig.~\ref{fig:Kendall}. 

Consider the set $\{\Pi_i\}$ of $n=60$ pairs $\Pi_i=(F_i,L_i)$ of mean blazar fluxes $F_i$ and corresponding flare likelihoods $L_i$. Construct various pairs of pairs $(\Pi_i,\Pi_j)$. A pair of pairs is called \textit{concordant} if either both $F_i<F_j$ and $L_i<L_j$, or both $F_i>F_j$ and $L_i>L_j$. It is called \textit{discordant} if either  $F_i<F_j$ but $L_i>L_j$, or  $F_i>F_j$ but $L_i<L_j$. The test statistics is defined as
$$
\tau=\frac{n_c-n_d}{\sqrt{\left( n_c+n_d+n_F \right) \left( n_c+n_d+n_L \right)}},
$$
where $n_c$ is the number of concordant pairs of pairs, $n_d$ is the number of discordant ones, $n_F$ and $n_L$ are the numbers of ties, or groups of repeated elements, in the $\{F_i\}$ and $\{L_i\}$ sets, correspondingly. 

% Don't change these lines
\bsp	% typesetting comment

\end{document}